# Experimental and Computational Studies of the Optical Properties of $CuAl_{1-x}Fe_xO_2$

Mina Aziziha[1], Saeed Akbarshahi[1], Suresh Pittala[2], Sayandeep Ghosh[3], Rishmali Sooriyagoda[1], Aldo H. Romero[1], Subhash Thota[3], Alan D. Bristow[1], Mohindar S. Seehra[1], and Matthew B. Johnson[1]*

*1. Department of Physics & Astronomy, West Virginia University, Morgantown, West Virginia 26506, USA*

*2. Department of Physics, Indian Institute of Science, Bangalore 560012, India*

*3. Department of Physics, Indian Institute of Technology, Guwahati, Assam 781039, India*

## Abstract

Delafossites are promising candidates for photocatalysis applications because of their chemical stability and absorption in the solar region of the electromagnetic spectrum. For example, $CuAlO_2$ has good chemical stability but has a large indirect bandgap (~3 eV), so that efforts to improve its absorption in the solar region through alloying are investigated. The effect of dilute alloying on the optical absorption of powdered $CuAl_{1-x}Fe_xO_2$ ($x$ = 0.0-1.0) is measured and compared to electronic band structures calculations using a generalized gradient approximation with Hubbard exchange-correlation parameter and spin. A new absorption feature is observed at 1.8 eV for $x$ = 0.01, which redshifts to 1.4 eV for $x$ = 0.10. This feature is associated with transitions from the L-point valence band maximum to the Fe-3$d$ state that appears below the conduction band of the spin-down band structure. The feature increases the optical absorption below the bandgap of pure $CuAlO_2$, making dilute $CuAl_{1-x}Fe_xO_2$ alloys better suited for solar photocatalysis.

**Introduction**

A potential photocatalyst should have high quantum-efficiency and be chemically stable. Ideally, they should be simple to synthesize and have a good surface-to-volume ratio. Metal oxides such as delafossite ($ABO_2$) crystals have attracted attention due to their chemical stability during electrolysis, [1–6] tendency for native *p*-type doping and often with small indirect gaps ($E_g <$ 1.5 eV) that makes them into good conducting electrodes. [7–9] Delafossite crystals, such as $CuAlO_2$, have wide direct optical bandgap ($E'_g = 3.8$ eV), exhibits a large exciton binding energy and have strong lattice polarization. [10–16] Moreover, in the powdered form, there are surface states that add to the optical spectra without necessarily offering useful charges for photocatalysis [14]. The optical absorption needs to be increased in the visible/solar range of the electromagnetic spectrum specifically to improve photocatalytic activity. As has been shown in $CuGaO_2$ [17], a promising way to do this is to alloy $CuAlO_2$ with smaller optical bandgap delafossites, such as $CuFeO_2$ ($E'_g \approx 2.1$ eV) [18]. In principle, dilute $CuAl_{1-x}Fe_xO_2$ should break symmetry of the pure $CuAlO_2$ sufficiently to lower the band-edge and create better photocatalytic absorption.

The calculation of the electronic band structure and the resultant optical absorption for pure non-alloyed delafossites is already challenging. Density functional theory (DFT), using local density (LDA) and generalized gradient (GGA) approximations, successfully predicts the crystal geometry of the insulators. However, they usually fail to predict the insulating band structure of transition-metal oxides such as $CuAlO_2$, [8,19–21] because they underestimate *d*-orbital correlations. [22] This can be overcome by including a Hubbard on-site Coulomb interaction (DFT+*U*) [23,24] or Green's function with Coulomb screening (GW), [12,15] both of which are more suitable for calculations of *d* orbitals. [25–27] Calculations using LDA/GGA+*U* correct the energy of the $d$ orbitals through the $U$ parameter, which is itself not self-interaction free. These oxides exhibit strong polaronic (large polaron constant: $\alpha_p$, $\alpha_p \sim 1$ [11]) and excitonic effects (exciton binding energy of ~0.5 eV [15]) that can be modeled in GW calculations with a many-body quasiparticle3 approach. Alternatively, hybrid functional calculations can mostly fix the self-interaction problem, improve the treatment of localized states and better predict the bandgap. [28] However, hybrid functionals are more computationally demanding than DFT+$U$ calculations for delafossites, and these demands are greater still when extending the calculations to dilute alloys that require large supercells not one unit cell. For example, one Fe atom in a 3×3×3 supercell is 4% Fe concentration ($CuAl_{0.96}Fe_{0.04}O_2$), whereas changes are expected observed for dilute concentrations at and below 1%. For practical reasons, taking a GGA+$U$ approach for dilute alloys of $CuAl_{1-x}Fe_xO_2$ will handle the charge distribution of the $d$ orbitals and make use of well-established values of $U$ [25] and can be applied across a wide concentration range.

In this paper, a comparison of optical absorption spectra and computed band structures are presented for the dilute alloys $CuAl_{1-x}Fe_xO_2$ ($x$ = 0.0-0.10) and $CuFeO_2$ ($x$ = 1.0). First-principle DFT calculations are performed to obtain the electronic band structure, and powdered alloys are measured in the photon energy range of 1 to 6 eV. Although powder samples exhibit pronounced defects, the consistent presence of the higher energy indirect and direct bandgaps associated with the $CuAlO_2$ are observed at 3.2 and 3.8 eV respectively. A new lower-energy absorption line is observed at about 1.8 eV that is associated with the Fe-3$d$ band, which appears within the calculated $CuAlO_2$ bandgap. This Fe-3$d$ absorption line is observed to redshift with increasing Fe concentration. However, both observed and supercell band structure results confirm the direct and indirect transitions do not change.

**Method**

**Experimental**

$CuAl_{1-x}Fe_xO_2$ powders ($x$ = 0.0, 0.01, 0.05, and 0.10) were prepared by the solid-state reaction method using $Cu_2O$, $Al_2O_3$, and $\alpha$-$Fe_2O_3$ as precursors in calculated stoichiometric ratios with heat treatments in static air at 1,100 °C [29]. The $x$ = 1.0 sample was synthesized in an argon ambient at 1,000 °C. Details of this work, including structural, chemical, IR/phonon, and magnetic characterization, are given in recent papers [30–33]. The optical properties of these alloy powders in UV–Visible–Near IR (UV-Vis-NIR) of 1 to 6 eV were measured at room temperature using diffuse reflectance spectroscopy employing a Shimadzu MPC3100 UV-visible spectrometer with an integrating sphere (using a $BaSO_4$ powder reference for background correction).

**Computational**

DFT calculations are performed using the Vienna Ab initio Simulation Package (VASP), [34–36] with the generalized gradient approximation (GGA) exchange-correlation [37] functional parameterized by Perdew, Burke and Ernzerhof (PBE) [26] and projector-augmented wave (PAW) pseudopotentials [27] with the valence-electron configurations $3p^63d^{10}4s^1$, $3p^63d^74s^1$, $3s^23p^1$, and $2s^22p^4$ for Cu, Fe, Al, and O, respectively. A simplified rotationally invariant PBE+$U$ approach [38] is taken with an on-site Coulomb parameter $U_{eff}$ = 4.0 eV, as suggested by Jain *et al.* [25], for both Fe and Cu. Spin polarization for Fe is also included. Structural relaxation is used, leaving the supercell volume, positions, and cell shape free to change. A 3×3×3 supercell is constructed from the rhombohedral structure (R$\bar{3}$m) primitive unit cell of the $CuAlO_2$ with 108 atoms to calculate comparable dilute alloys. There are many possible configurations for Fe substituting of Al, but most of them can be ruled out by symmetry using a site-occupation disorder (SOD) code [39]. The structural relaxation of the resulting disordered supercell uses a 4×4×4, Γ-centered, *k*-point grid, and for all calculations, the electron wave functions were expanded in a plane-wave basis set with cut-off kinetic energy set to 600 eV. The Fold2Bloch code [40,41] was employed to relate the supercell band structure to the primitive basis representation using Bloch-spectral-density.

**Results and Discussion**

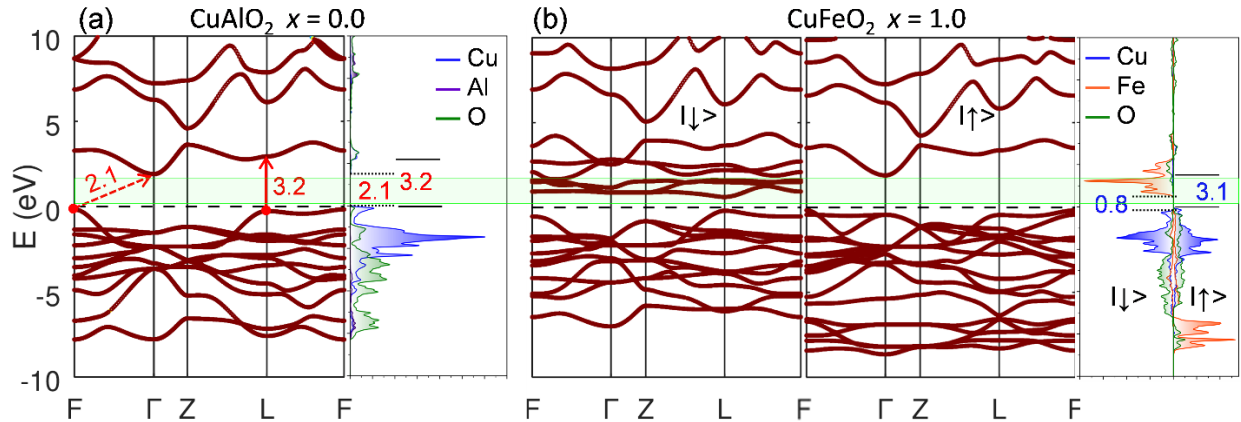

Figure 1: (a) Electronic band structure of $CuAlO_2$ with the density of states, adjacent. (b) Electronic band structure of the $CuFeO_2$ spin down ($|\downarrow\rangle$) on left spin up ($|\uparrow\rangle$) on right, as labeled, with density of states, adjacent. The black dashed lines show the Fermi energy. A shaded rectangle highlights the energy region where there are no states in $CuAlO_2$ and closely spaced states in $CuFeO_2$.

Figure 1(a) shows the computed band structure and density of states (DOS) for $CuAlO_2$. The lowest conduction band (CB) consists of O-2$p$ orbitals with a minimum located at $\Gamma$-point. The valence band (VB) consists of Cu-3$d$ orbitals with maxima located at the L- and F-points. A direct optical bandgap of 3.2 eV is marked on the figure at the L-point, which is slightly lower, but consistent with other reported values (3.4-4.0 eV). [8,13,24,42,43] Similarly, a smaller indirect fundamental bandgap ($E_g$ = 2.1 eV) is marked on the figure, between the VB maximum at the F-point and the CB minimum at $\Gamma$-point. This is slightly higher than earlier LDA/GGA calculations that calculate indirect bandgaps between 1.6 and 2.0 eV, [7–9], but these earlier calculations without the addition of the Hubbard parameter are known to under-estimate $E_g$. Any correspondence between such calculations and optical features observed below 2 eV is a misattributing of the bandgap to Cu vacancies and Al anti-site defect states in real crystals [10–12].

Figure 1(b) shows the band structure for spin-up $|\uparrow\rangle$ and spin-down $|\downarrow\rangle$ states separately along with the associated spin-polarized DOS CuFeO$_2$. Both spin-polarizations have VB states that consist of Cu-3$d$ orbitals. The VB maximum is located at the F-point in the Brillouin zone with a near-maximum at the L-point. By contrast, spin-polarization has a significant effect on the CB with $|\downarrow\rangle$ and $|\uparrow\rangle$ states consisting of Fe-3$d$ and O-2$p$ orbitals and with minima at the L- and Γ-points respectively. Consequently, CuFeO$_2$ has spin-polarized bandgaps, with an indirect $E_g = 0.8$ eV to $|\downarrow\rangle$-states and optically-allowed bandgap at $E_g' = 3.1$ eV to $|\uparrow\rangle$-states. These values are consistent with reported observations of $E_g = 1.1$ eV and two direct optical bandgaps at $E_g' = 2.0$ and 3.4 eV respectively. [44–46] Exchanging Al for Fe significantly narrows the fundamental gap and hence is expected to change the optical properties in the visible and near-infrared. The shaded rectangle running horizontally across Fig. 1 between energies 0.2 and 2 eV serves to accentuate the lack of calculated states in CuAlO$_2$ and the closely spaced bands present for CuFeO$_2$.

Figure 2 shows the GGA+$U$ first-principle supercell calculations of the CuAl$_{1-x}$Fe$_x$O$_2$ for $x$ = 0.04, 0.08 and 0.11. The top and bottom rows show the $|\downarrow\rangle$ and $|\uparrow\rangle$ band structures respectively, with the associated DOS plotted alongside. For comparison, the DOS for $x = 0$ and 1 are included (without their band structures) on the upper left and lower right for comparison across the entire Fe concentration range. The color spectrum (from white to dark red) is associated with the spectral weight from 0.0 to 1.0, which denotes both the magnitude of the corresponding Bloch character and degeneracy of the bands at any point in $k$-space when unfolding the supercell band structure onto the effective unit cell of the ternary compound [47].

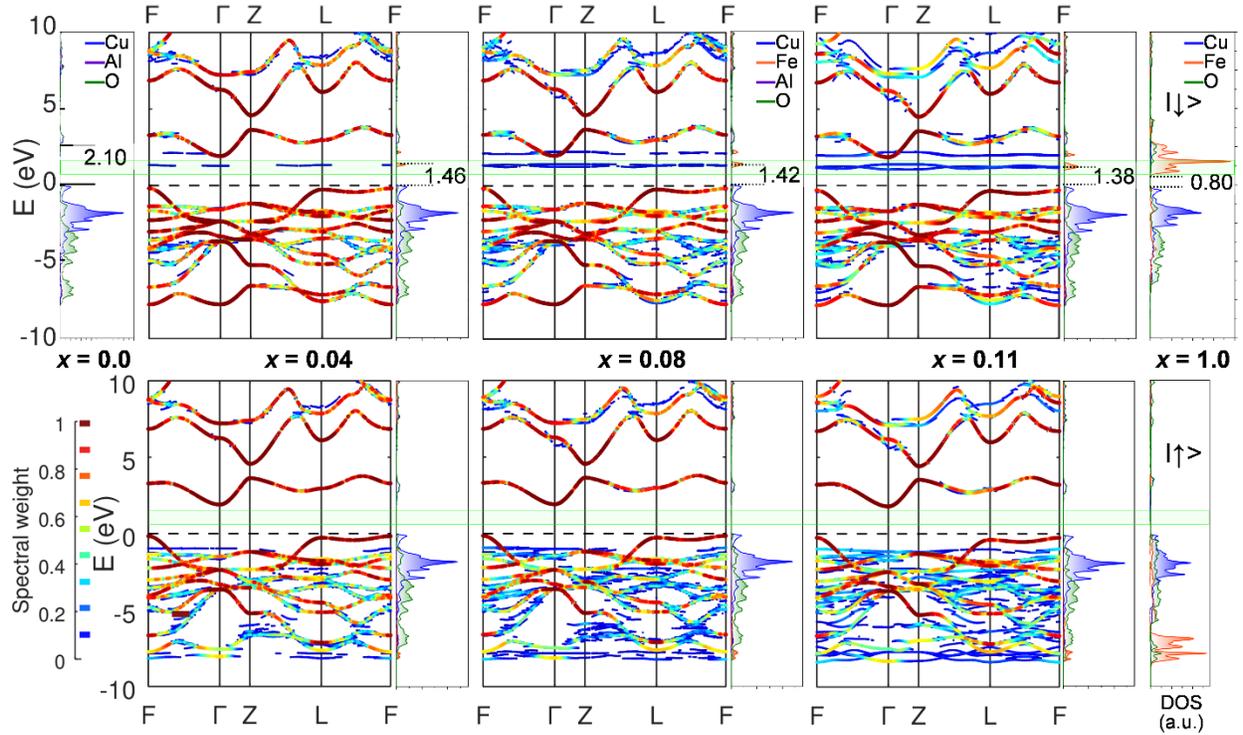

Figure 2: shows the spin-polarized electronic band structures of $CuAl_{1-x}Fe_xO_2$ for $x = 0.0, 0.04, 0.08, 0.10,$ and $1.0$ and the associated density of states adjacent, calculated for the supercell made of $3\times3\times3$ primitive unit cell. The arrows show the spin state of the density of states. The spectral weight scale of the colors are shown in bottom left. The black dashed lines show the Fermi energy. Shaded rectangles highlight the energy region where the Fe-3d related state builds.

In dilute $CuAl_{1-x}Fe_xO_2$, two changes may be expected: (1) Due to breaking the inversion symmetry in $R\bar{3}m$ structure, the forbidden direct transition at Γ-point becomes allowed. However, these calculations clarify that the minimum optical transition is at the L-point, which indicates that this change should not be significant. (2) Spin-down Fe-3$d$ states are added into the $CuAlO_2$ bandgap, below the CB minimum, and can decrease the fundamental bandgap. Spin-up Fe-3$d$ orbitals are added to the VB but below VB maximum without changing the bandgap. Indeed, the increase in $|\downarrow\rangle$ Fe-3$d$ orbitals do narrow the gap, but these orbitals are not well defined in $k$-space, so they are weak and hence have a low spectral weight (indicated by blue). Nonetheless, increasing Fe concentration increases the coverage of the blue bands throughout the band structure. Meanwhile, $|\uparrow\rangle$ Fe-3$d$ orbitals have increased the coverage of blue bands in the VB with minimal modification to the CB band structure. Overall, the addition of band compared to CuFeO$_2$ is expected to increase the optical absorption in the visible and near-infrared. The shaded rectangle running horizontally across Fig. 2 between energies 0.2 and 2 eV serves to accentuate the lack of states in the calculated CuAlO$_2$ and this buildup of states in the CuAl$_{1-x}$Fe$_x$O$_2$ alloys.

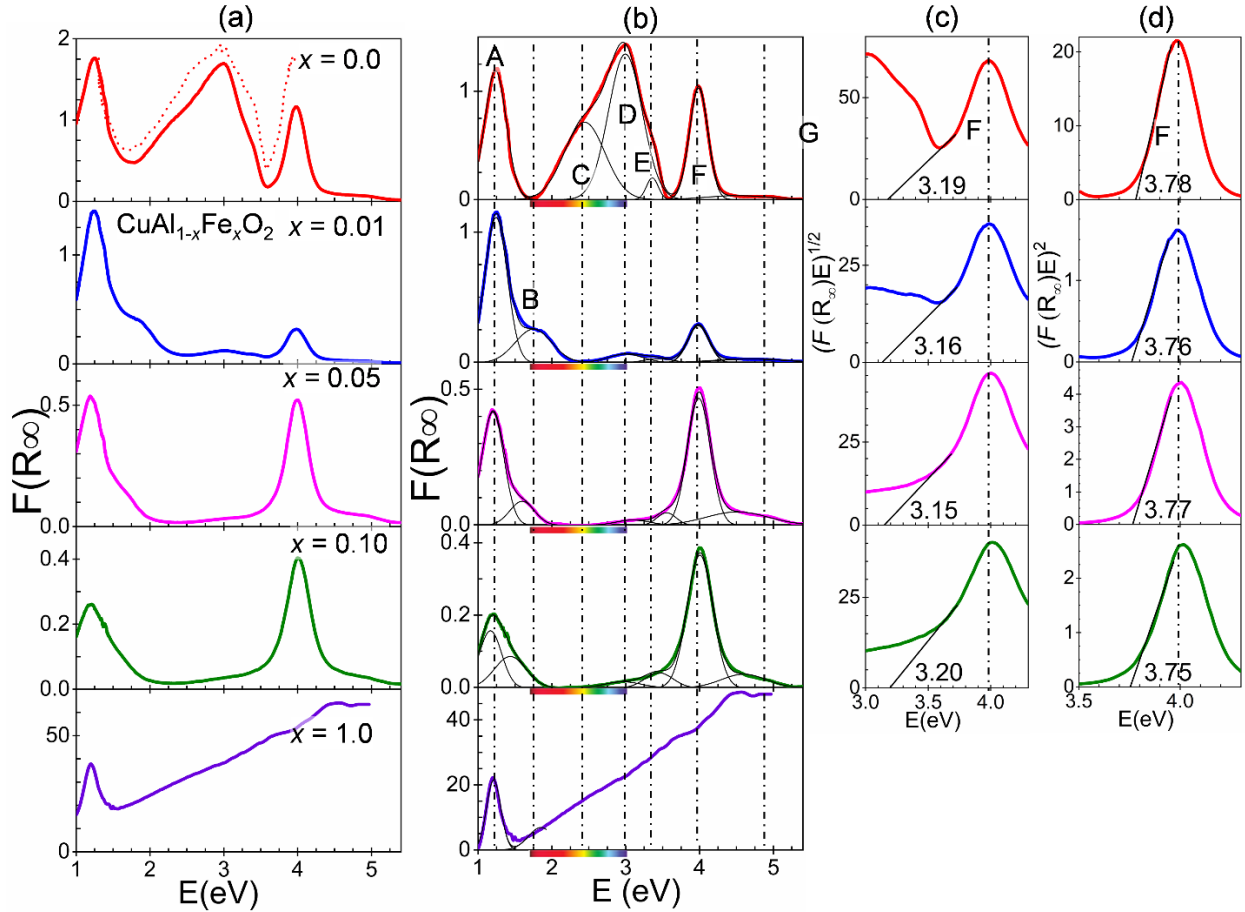

Figure 3: (a) Raw F($R_\infty$) calculated from reflection data by Kobelka-Monk theory plotted for CuAl$_{1-x}$Fe$_x$O$_2$, $x$ = 0-1, the subtracted background is shown by dashed lines. For x = 0.0 we include data for a recently synthesized powder [50] (dotted) (b) F($R_\infty$) without the background and the fitted peaks using Gaussians functions, shown as dashed lines and labeled A–G. (c) and (d) are Tauc plots for the direct and indirect bandgaps, respectively, with the fitted lines and intercepts shown.

Room-temperature optical reflection ($R$) spectra were converted using the Kubelka-Munk method for highly scattering media [48,49] into the ratio of the absorption coefficient ($\alpha$) and scattering coefficient ($S$) using $\alpha/S = F(R_\infty) = (1 - R)^2/2R$. Figure 3(a) shows F($R_\infty$) for powdered $CuAl_{1-x}Fe_xO_2$ with $x$ = 0.0-1.0. Typically, such data have an energy-dependent background that can be seen as lifting the troughs in the spectra above $F(R_\infty) = 0$. Figure 3(b) shows $F(R_\infty)$ with the background subtracted and deconvolution of the peaks with a series of Gaussians. The peaks of each feature are labeled on the $CuAlO_2$ spectra and drawn below the spectra for comparison. In the spectra, except for $x$ = 1.0, there are multiple peaks labeled A–G with increasing photon energy. The spectrum for $CuFeO_2$ ($x$ = 1.0) is distinctly different, showing a monotonic increase above peak A, with some structure above 4 eV. This increase is consistent with other studies [50,51] and is a reflection of the closely spaced bands that are present in the DFT calculations (see Fig. 1).

Powder samples are known to exhibit impurities, defects, and surface states in the optical absorption that cannot be compared directly to calculations based on single crystals without intrinsic or extrinsic defects. Previously, defect absorption peaks [14,52] have been identified and can be related to the large absorption peaks labeled A, C, D, and E. X-ray photoelectron spectroscopy indicate that these powders have a significant amount of $Cu^{2+}$ on their surfaces [30]. With that in mind, peak A at ~1.2 eV is likely due to $Cu^{2+}/Cu^+$ intervalence transitions associated with mixed oxidation states [53]. This is consistent with the feature decreasing in strength with the addition of a little Fe. Peaks C, D, and E are probably due to oxygen vacancies and oxygen interstitials, which also decrease in strength with Fe concentration. [14] The weak high-energy feature G is probably associated with a transition to one of the higher energy bands and not important here.

Peak F at ~4.0 eV has been identified as the direct optical bandgap ($E_g'$) at the L-valley. [16,52] Figure 3(c) and (d) show Tauc plots to determine direct and indirect ($E_g$) for $x$ = 0.0, 0.01, 0.05, and 0.10. In the Tauc method [54,55], absorption is associated with the band edge such that $[F(R_\infty)\hbar\omega]^n \propto (\hbar\omega - E_g^{(\prime)})$, where $\hbar$ is the reduced Plank's constant, $\omega$ is the optical frequency, and the exponent is set as $n = 2$ or ½ for direct or indirect transitions respectively [56,57]. From the measurements, the fundamental gap corresponding to the indirect F-Γ transition is ~3.8 eV, and the direct optical gap ($E_g'$) corresponding to L-valley transition is ~3.2 eV. Both values are consistent with the calculated range of fundamental and optical bandgaps, $E_g$ and $E_g'$ respectively.

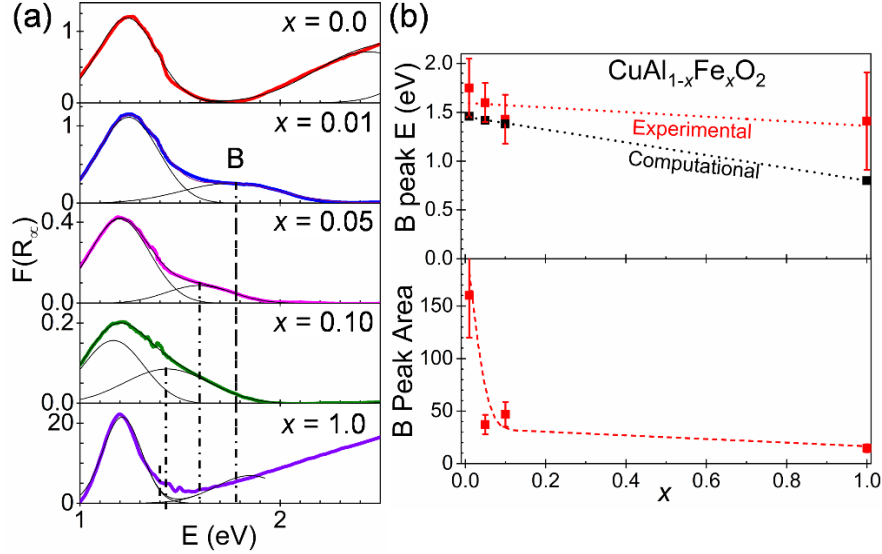

Figure 4: (a) Expanded view of Fe-related absorption peak (B). (b) Top: Plot of the energy of the Fe-related absorption (B) and position of the Fe-related state in the modeled band structure *vs. x*. (b) Bottom: plot of area under the peak for the Fe-related absorption (B). The dashed lines are the guides to the eye to follow the trend.

The clear positive spectral indicator for the introduction of Fe into the alloy is peak B. It emerges as a shoulder on the high-energy edge of peak A in the $x = 0.01$ spectrum and remains in the $x = 0.05$ and 0.10 spectra as well. By comparison, peaks C and D are suppressed with the introduction of Fe, and the defect-related peak A and the band-edge-related peak F are both observed for the series of dilute alloys. This result is unsurprising for peak F because even at $x = 0.1$, Vegard's law predicts bandgaps closer to pure $CuAlO_2$. [58] Peak B is central to this work and is the highlight in Figure 4(a) with a close up of the relevant spectral range.

The position of peak B in the F($R_\infty$) spectrum corresponds to the appearance of the $|\downarrow\rangle$ Fe-3$d$ bands below the CB minimum in the DFT calculations in the bandgap (see Fig. 2). Experimentally, peak B is centered at 1.8 eV in the $x$ = 0.01 spectrum and redshifts to 1.4 eV in the $x$ = 0.10 spectrum. This redshift with Fe concentration is not observed for any other peaks shown in Fig. 3(b) and can be compared to the position of the Fe-3$d$ bands as a function of $x$ in Fig. 2. In Fig. 2, transitions between the VB maximum to the Fe-3$d$ bands are labeled for the series of dilute alloys, also following a redshift of the band with increasing Fe concentration. Summary of the observed peak center energies and computed transition energies are shown in the top panel of Fig. 4(b). The error bars are half the full width at half maximum of peak B. The experimental (squares) and computed (circles) are fitted to linear lines (dotted). The trends of experiment and computation show good agreement and that peak B depends on the introduction of Fe and the concentration of Fe. To the best of the authors' knowledge, this is the first observation of a purely Fe-dependent mode observed in the CuFe$_x$Al$_{1-x}$O$_2$ crystal.

Finally, in the bottom panel of Fig. 4(b) we plot the area (height times full width half maximum) of peak B as a function of Fe concentration. It is seen that the peak is non-existent for CuAlO$_2$, is strong for $x$ = 0.01, and generally decreases with increasing Fe concentration. The peak area decrease is slightly non-monotonic, dipping down at $x$ = 0.05 and slightly recovering at $x$ = 0.10, before continuing to decrease to its ultimate area value in CuFeO$_2$. The dip at $x$ = 0.05 is reminiscent of observations in delafossite CuGa$_{1-x}$Fe$_x$O$_2$, where Fe is ordered in planes as the number of Fe atoms is commensurate with a large supercell and produces an additional degree of long-range order. [17] The overall decrease in the area of peak B is somewhat surprising given the increase in Fe and the build-up of the Fe-3$d$ states (shown in Fig. 2). This may be the result of powder samples, with large surface-to-volume ratios and related defects, or the ordering of Fe with increased Fe concentration.

**Conclusions**

The optical reflection spectra are measured for $CuAl_{1-x}Fe_xO_2$ for $x = 0.0-1.0$ powders, and band structures are calculated by the first-principle DFT. A PBE+$U$ approach is used for the calculations, and $U$ is optimized to best match with the experimental results. According to the band structure calculations, the fundamental indirect bandgap is F-Γ transition, experimentally observed at ~3.2 eV, and the direct optical bandgap is at the L-point, observed at ~3.8 eV. The positions of these bandgaps are insensitive to Fe content ($x = 0.01-0.10$) in agreement with the computed band structures.

Fe concentration does result in a new optical feature near ~1.4 eV that undergoes a redshift with increasing concentration and is associated with Fe-3$d$ orbitals that form a band below the $CuAlO_2$ conduction band edge. Earlier work suggested alloying breaks the inversion symmetry in $R\bar{3}m$ structure, such that the otherwise-forbidden direct transition at Γ-point becomes allowed. However, the presented DFT calculations confirm that the minimum direct bandgap remains at L-point, ruling out that possibility. Nonetheless, the Fe-concentration-dependent optical feature can be tuned by composition and offers a new absorption mechanism near 1.6 eV in $CuAl_{1-x}Fe_xO_2$. This result may make powders of dilute-Fe delafossite alloys more appealing for applications such as photocatalytic activity.